\documentclass[amssymb,useAMS,prd,aps,amsmath,twocolumn,superscriptaddress,nofootinbib]{revtex4}
\usepackage{color}
\usepackage{pslatex}
\usepackage{graphicx}

\def\lsp{\tilde\chi^0_1}
\def\ra{\rightarrow}

\begin{document}

\title{Can neutralinos in the MSSM and NMSSM scenarios still be light?}

\author{Daniel Albornoz V\'asquez}
\affiliation{Astrophysics department, DWB building, Keble Road, OX1 3RH Oxford}
\affiliation{LAPTH, U. de Savoie, CNRS,  BP 110,
  74941 Annecy-Le-Vieux, France.}

\author{Genevi\`eve B\'elanger}
\affiliation{LAPTH, U. de Savoie, CNRS,  BP 110,
  74941 Annecy-Le-Vieux, France.}

\author{C\'eline B\oe hm}
\affiliation{LAPTH, U. de Savoie, CNRS,  BP 110,
  74941 Annecy-Le-Vieux, France.}

\author{Alexander Pukhov}
\affiliation{Skobeltsyn Inst. of Nuclear Physics, Moscow State Univ., Moscow 119992, Russia}

\author{Joseph Silk}
\affiliation{Astrophysics department, DWB building, Keble Road, OX1 3RH Oxford, England}

\date{today}

\begin{abstract}
Since the recent results of direct detection experiments at low mass, many authors have revisited the case 
of light (1 -10) GeV WIMPs. In particular, there have been a few attempts to explain the results from the 
DAMA/LIBRA, CDMS and/or CoGeNT experiments  by invoking neutralinos lighter than 15 GeV. Here we show that in the MSSM,  
such light particles  are completely ruled out by the TEVATRON limits on the mass of the pseudoscalar Higgs.  
On the contrary,  in the NMSSM, we find that light neutralinos could still be viable 
candidates. In fact, in some cases, they may even have an elastic scattering cross section on nucleons in 
the range that is needed to explain either the DAMA/LIBRA, CoGeNT or CDMS recent results. 
Finally, we revisit the lowest limit on the neutralino mass in the MSSM and find that neutralinos 
 should be heavier than $\sim$ 28 GeV  to evade present experimental bounds.
\end{abstract}
\maketitle

\section{Introduction}
Recently, both the CDMS  and CoGeNT experiments have announced results which have been regarded as possible hints of the existence of dark 
matter particles. In particular, the CDMS experiment has claimed the detection of two ``anomalous'' events in a blind analysis \cite{Ahmed:2009zw}  
while the CoGeNT experiment has observed an unexplained rise in a p-type point contact (PPC) spectrum devoid of most surface 
events \cite{Aalseth:2010vx}. Together, these results might confirm the long-standing claims of detection of dark matter annual modulation by  the 
DAMA/LIBRA experiment. Therefore, these findings deserve some attention. 
  
The most puzzling aspect of these recent claims is that they seem compatible with the existence of rather light dark matter particles.  
For example,
the assumption of dark matter particles with a mass between 7 and 11 GeV  seems to provide a good fit to CoGeNT data (even though the
null hypothesis also provides a similar fit) while particles from 7 GeV to 40 GeV could explain the CDMS events. This mass range is also compatible 
with the DAMA/LIBRA \cite{Bernabei:2008yi,Bernabei:2010mq} findings which favour particles in the 10-15 GeV range assuming spin-independent (SI)
interactions and no channeling.

Although these claims sound quite exciting, they raise several issues.
First of all, CDMS  claimed detection of two events only. This is not statistically significant to enable any firm conclusion related to dark matter. 
However, detection of more than five events would have been in conflict with the negative results of XENON 10 \cite{Aprile:2010bt}. 
Secondly, the possible interpretation of a $\sim 7-11$ GeV dark matter candidate with a SI cross section around $10^{-41}-10^{-40}{\rm cm}^2$(as proposed 
by the CoGeNT collaboration) is 
severely challenged by the negative results of the XENON 100 experiment \cite{Aprile:2010um}.  Nevertheless, it was also demonstrated that 
the uncertainties on the dark matter escape velocity $v_{esc}$ and scintillation efficiency $L_{eff}$ may offer a way to reconcile 
both XENON 100 and CoGeNT results~\cite{Savage:2010tg,Manzur:2009hp,Collar:2010nx,Collar:2010gd,Collaboration:2010er,Collar:2010gg,McCabe:2010zh,Gastler:2010sc}. 
Finally, the CDMS and  CoGeNT events seem to exclude the high value of the spin-(in)dependent cross section that is favoured by the DAMA/LIBRA experiment.

Nevertheless, it is worth mentioning that the mass range that is under consideration is not excluded neither by the ISR constraints at LEP \cite{Belanger:2003wb} 
nor by the potential Sunyaev-Zel'dovich effect that might be generated by relatively light dark matter particles in clusters of galaxies. \cite{Lavalle:2009fu,Boehm:2008nj}.
Although it is premature to draw any conclusion about the nature of the DAMA/LIBRA,  CDMS and CoGeNT results, it is certainly worth reinvestigating the possible dark matter candidates that are expected in the low mass region. 

In this paper, we shall only consider 
supersymmetric candidates  and in fact focus more specifically on the neutralino ($\tilde{\chi}_0$). We are seeking a 
light $\tilde{\chi}_0$ with a relatively high elastic scattering cross-section with matter. In order to retain the 
connection with the ``dark matter'' interpretation, one should also require that this candidate has a relic density which 
represents a subsequent fraction (if not all) of the dark matter abundance inferred from the combination of CMB data with 
other cosmological (e.g. SN, BAO) observations \cite{Bennett:2010jb}.  

Recently, the authors of Ref.~\cite{Bottino:2009km} claimed that light neutralinos 
in MSSM-EWSB scenarios with non-universal gaugino masses could actually do the job. 
This work was followed by  other claims (using supersymmetric extensions of the 
Standard Model
\cite{Kuflik:2010ah,Fitzpatrick:2010em,Bae:2010hr,Hisano:2009xv,Asano:2009kj,Gogoladze:2009mc}) 
to explain CoGeNT and/or CDMS events. 
However, light MSSM neutralinos  were in fact investigated 
before these experimental claims \cite{Belanger:2003wb,Hooper:2002nq,Bottino:2002ry,Bottino:2009km,Hooper:2008au} and 
just after the study of the astrophysical signatures that are to be expected from light WIMPs \cite{bens}. 
Neutralino masses down to $\sim$ 5 GeV were found. However in Ref.~\cite{Feldman:2010ke}, it was pointed out 
that the improved measurement of $B_s \rightarrow \mu \mu $ excludes such low neutralino masses. Very light neutralinos that
would constitute hot dark matter were however found to be consistent with all experimental constraints~\cite{Dreiner:2009ic}.

Here, we reinvestigate the MSSM in light of  
the latest TEVATRON results to determine whether light neutralinos can explain the direct detection signals or not.
We find that, when taking into account the Tevatron results on the supersymmetric Higgs at large values of
$\tan\beta$, the lower limit on the mass of the neutralino increases to 28~GeV.  We then consider an extension of the 
MSSM with an extra singlet, the NMSSM. Neutralino dark matter in singlet extensions of the MSSM were first investigated in 
Ref.~\cite{Abel:1992ts}. When the singlet Higgses are light we find scenarios with neutralinos of a few GeV that evade all constraints
and yet predict a direct detection signal in the region preferred by recent experiments.

In section II, we describe the method used for exploring the parameter space. We analyse scenarios with light neutralinos in  the MSSM assuming different priors in section III   
while  the results for the NMSSM are presented in section IV.

\section{Method \label{sec:Meth}}
To  efficiently explore the multi-dimensional parameter space,  we have performed a Markov Chain Monte Carlo analysis 
(based on the Metropolis-Hastings algorithm). We have used micrOMEGAs2.4 \cite{Belanger:2006is,Belanger:2010pz} 
to compute all observables. This code
relies  in turn   on SuSpect~\cite{Djouadi:2002ze}  for calculating the particle spectrum in the MSSM  and on NMSSMTools~\cite{Ellwanger:2005dv} for calculating the particle spectrum and the various collider  and B-physics constraints in the NMSSM.

We use the method of burn-in chains, i.e. we first explore the parameter space till we find a point with a non-vanishing likelihood. When such a point is found, we continue the chains, keeping all the points that are retained by the MCMC. However, since it is difficult and time-consuming to find a 
good starting point, we require to speed up the process that the likelihood times the prior (hereafter referred to as $Q$) associated with the starting point exceeds the value $Q> 10^{-12}$, and use an exponential
 prior on $m_{\chi}$ to make sure that the starting point is within close proximity of the low neutralino mass region. However, when this point is found, we replace the exponential prior on $m_{\chi}$ by a flat prior. Since low mass neutralinos are quite unlikely with respect 
 to heavier ones (and since finding them also requires a certain amount of fine-tuning), we have decided to perform two independent scans. One aims at exploring the mass region ranging from 0 to 15 GeV (this range of mass is particularly relevant for the CoGeNT and DAMA/LIBRA interpretation) and the second one aims at exploring the range from 0 to 50 GeV (to include the preferred region of the two WIMP recoil-like events reported by 
CDMS-II). A proper exploration of the parameter space is obtained after generating approximately 50 chains of $10^5$ points each.

The total likelihood function for each point is the product of  the likelihood functions evaluating the goodness-of-fit to all the data set that are displayed in 
Table.~\ref{likelihoods}. These include B physics observables, 
the anomalous magnetic moment of the muon, $(g-2)_{\mu}$, the Higgs and sparticles masses obtained from LEP 
and the corrections to the $\rho$ parameter. 
For the MSSM case, only LEP mass limits on new particles were taken as a sharp discriminating criterion with ${\cal L}= 0 \;{\rm or}\; 1$. 
Other criteria had some tolerance. For the NMSSM, limits on the Higgs sector, on the Z partial width and on neutralino production as computed by NMSSMTools
were also taken  as a sharp discriminating criterion.

We use a Gaussian distribution for all observables with a 
preferred value $\mu\pm\sigma$,
\begin{equation}
F_2\left(x,\mu,\sigma\right) = e^{-\frac{\left(x-\mu\right)^2}{2\sigma ^2}}
\end{equation}
and
\begin{align}
F_3\left(x,\mu,\sigma\right) = \frac{1}{1+e^{-\frac{x-\mu}{\sigma}}}.
\end{align}
for observables which only have lower or upper bounds. 
The tolerance, $\sigma$, is negative (positive) when one deals with an upper (lower) bound.

\begin{table}[h]
\caption{List of constraints, from Ref.~\cite{Nakamura:2010zzi} unless noted otherwise}
  \centering
    \begin{tabular}{|c|c|c|c|}
    \hline
&&&\\
\rm{constraint} \ &\rm{value/range} \ & \rm{tolerance} \ & \rm{applied} \\
&&&\\
\hline
Smasses \ & - \ & none \ & \rm{both} \\
\hline
$\Omega_{WMAP} h^2$ \ &0.01131 - 0.1131 \ & 0.0034 \ & both \\
\hline
$(g-2)_{\mu}$ &$25.5 \ 10^{-10}$  & stat: $6.3 \ 10^{-10}$ \ & both \\
&&sys: $4.9 \ 10^{-10}$ \ & \\
\hline
$\Delta \rho$ & $\leq 0.002$ &0.0001 \ & MSSM \\
\hline
$b \rightarrow s \gamma$ & $3.52 \ 10^{-4}$~\cite{Barberio:2008fa,Misiak:2006zs}& th: $0.24 \ 10^{-4}$ \ & \rm{both} \\
&& exp: $0.23 \ 10^{-4}$ \ & \\
\hline
$B_s \rightarrow \mu ^+ \mu ^-$ & $\leq 4.7 \ 10^{-8}$ & $4.7 \ 10^{-10}$ \ & both  \\
\hline
$R(B \rightarrow \tau \nu)$ &1.28~\cite{Barberio:2008fa} &0.38 \ & \rm{both} \\
\hline
$m_H$ & $\geq 114.4$& 1\% \ & \rm{MSSM}  \\
\hline
$Z \rightarrow \chi_1 \chi_1$ & $\leq 1.7$ MeV& 0.3 MeV \ & \rm{MSSM}  \\
& & \rm{none} \ & \rm{NMSSM}  \\
\hline
$e ^+ e ^- \rightarrow \chi_1 \chi_{2,3}$  & $\leq 0.1$ pb ~\cite{Abbiendi:2003sc} & 0.001 pb \ & \rm{MSSM} \\
&  & \rm{none} \ & \rm{NMSSM} \\
\hline
$\Delta M _s$ & $117.0 \ 10^{-13}$  GeV & th: $21.1 \ 10^{-13}$ GeV \ & \rm{NMSSM} \\
&& exp: $0.8 \ 10^{-13}$ GeV \ & \\
\hline
$\Delta M _d$ & $3.337 \ 10^{-13}$ GeV  & th: $1.251 \ 10^{-13}$ GeV\ & \rm{NMSSM}  \\
&& exp: $0.033 \ 10^{-13}$ GeV \ & \\
\hline
\end{tabular}
    \label{likelihoods}
\end{table}

Finally we also require that the neutralino relic density satisfies 
\begin{equation} 100 \% \ \Omega_{WMAP} h^2> \Omega_{\chi} h^2> 10 \% \  \Omega_{WMAP} h^2,
\label{wmap}
\end{equation}
with $\Omega_{WMAP} h^2=0.1131 \pm 0.0034 $ \cite{Komatsu:2008hk}. 
The cases where  $\Omega_{\chi}< \Omega_{WMAP}$ should correspond to scenarios in which there is either another  (if not several)
type of dark matter particles in the galactic halo \cite{Boehm:2003ha} or a modification of gravity (cf e.g. \cite{Skordis:2005xk}). In case 
of a multi component dark matter scenario, there could be either very light e.g.\cite{Boehm:2002yz,Boehm:2003bt,Boyarsky:2009ix,Gelmini:2009xd} or very heavy  particles (including very heavy neutralinos), depending on the findings of direct detection experiments.

\section{MSSM scenarios \label{sec:EWSB}}

In what follows, we consider the MSSM  with input parameters defined at the weak scale. We assume
minimal flavour violation and equality of the soft masses between sfermion generations. 
We further assume a common mass $m_{\tilde{l}}$ for all sleptons, 
and for all squarks $m_{\tilde{q}}$ (but we have checked that we found consistent results by relaxing this universality assumption).  
We allow for only one non-zero trilinear coupling, $A_t$. 
The gaugino masses $M_1$ and $M_2$ are free parameters which, in
particular, allows to have $M_1\ll M_2$ implying a light neutralino much below the EW scale. 
The parameter $M_3$ satisfies the usual universality condition in GUT scale model, that is $M_3=3M_2$. 
The Higgs bilinear term, $\mu$, the ratio of Higgs vev's, $\tan\beta$ and the pseudoscalar mass $M_A$ are also free parameters.
  
This MSSM-EWSB model with only eight parameters can reproduce the salient features of neutralino dark matter. 
Indeed, apart from the mass of the LSP, the most important parameters are 
the gaugino/higgsino content of the LSP, determined by $\mu$ and $M_1$, $M_2$, $\tan \beta$, 
as well as the mass of the pseudoscalar which can enhance
significantly neutralino annihilations into fermion pairs.

\begin{figure*}[bt]
	\centering	\includegraphics[width=15cm]{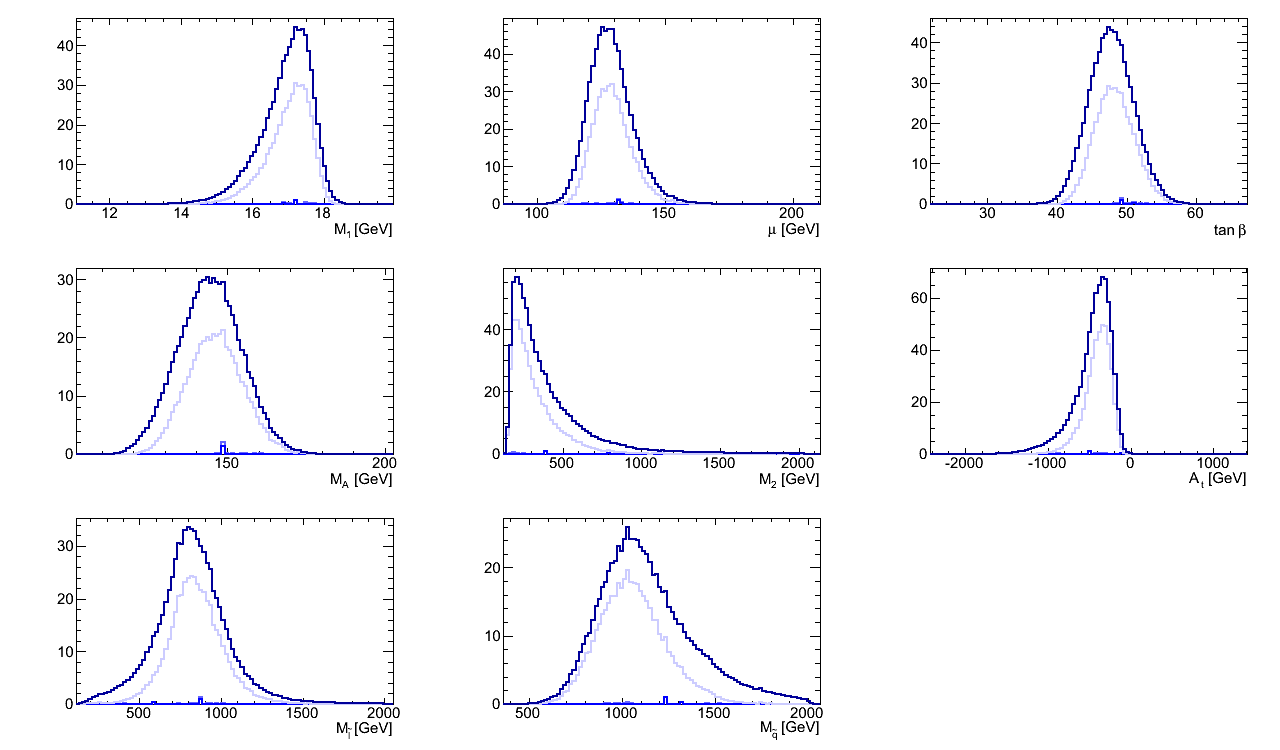} 
	\caption{MSSM-EWSB scenario for $\mu>0$ and $m_{\chi} < 15$ GeV. These plots represent the rescaled weight 
	$Q/Q_{max}$  of the points selected by the MCMC versus
	  the free parameters that we have considered. Curves in dark blue correspond to points 
	  with a likelihood greater than 99.4 $\%$; Curves in blue, correspond to points with 
	likelihood greater than 95.4 $\%$ and smaller than 99.4 $\%$ of the maximum Likelihood and  points
	 in pale blue are all the remaining points having a likelihood greater than 68 $\%$.}
	\label{fig:param}
\end{figure*}

\subsection{Neutralino masses smaller than 15 GeV}

\begin{figure}[h]
	\centering \includegraphics[width=8cm]{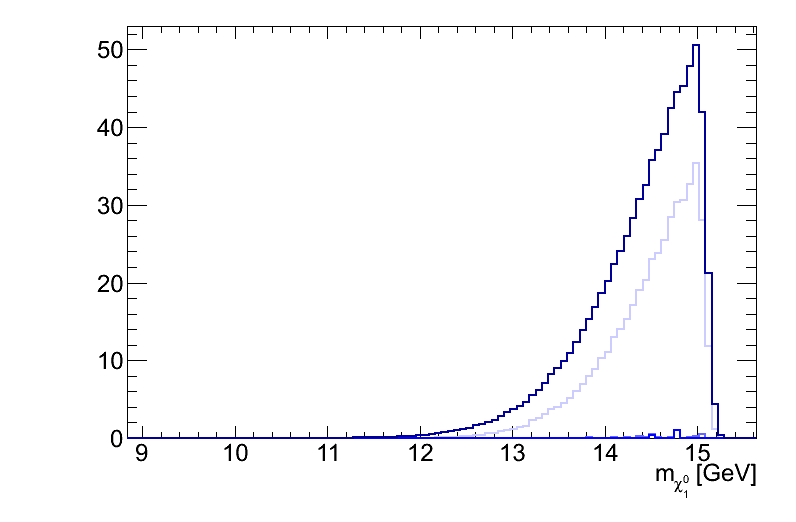}
	\caption{$Q/Q_{max}$ with respect to the neutralino mass in
	MSSM-EWSB scenario for $\mu>0$ and $m_{\chi} < 15$ GeV.	We use the same color code as in Fig.~\ref{fig:param}.}
	\label{fig:distrib}
\end{figure}

To sample the low neutralino mass range, we take our priors in the range 
\begin{equation}
\begin{array}{cc}
M_1 \in [1,100] {\rm GeV}  & M_2 \in [100, 2000]{\rm GeV} \nonumber\\
\mu \in [0.5,1000] {\rm GeV}  &\tan \beta \in  [1,75]  \nonumber\\
m_{\tilde{l}} \in [100,2000]{\rm GeV} & m_{\tilde{q}} \in  [300,2000]{\rm GeV} \nonumber\\
A_t \in [-3000,3000]{\rm GeV} & m_A \in  [100,1000] {\rm GeV} 
\end{array}
\end{equation}
We consider separately the cases   $\mu>0$ and $\mu<0$.
The results of our MCMC simulations for $\mu>0$ are displayed in  Fig.~\ref{fig:param} and ~\ref{fig:distrib}.

Fig.~\ref{fig:param} represents $Q/Q_{max}$,  the weigth normalized to the best weight,
with respect to the free parameters that we have considered. The first plot shows that the bino mass is peaked
 around $M_1 \in [15,19]$ GeV while the second plot shows that $\mu$ is below 150 GeV. That is, it is near the lower 
 bound that satisfies the LEP limits on charginos. 
Thus, the LSP is dominantly bino-like with a small Higgsino component.  

The third and fourth plots in Fig.~\ref{fig:param} show that $\tan \beta$ is very large ($\tan \beta \in [40,60]$) and 
$m_A$ is relatively small ($m_A \in [120,170]$ GeV). 
This basically indicates that the main neutralino pair annihilation proceeds through the s-channel exchange of a light pseudo-scalar Higgs boson. 
The results also show that the sleptons and squarks are preferably heavy ($m_{\tilde{l}} \in [500,1200]$ GeV and $m_{\tilde{q}} \in [0.8,2]$ TeV). 

In Fig.~\ref{fig:distrib}, we display the same quantity but with respect to the neutralino mass. As one can see, the preferred 
value for the neutralino mass $m_{\chi}$ lies in  between 13 and 15 GeV. We found no scenario where the neutralino mass would be smaller than 10 GeV.

\begin{figure}[h]
\centering \includegraphics[width=8cm]{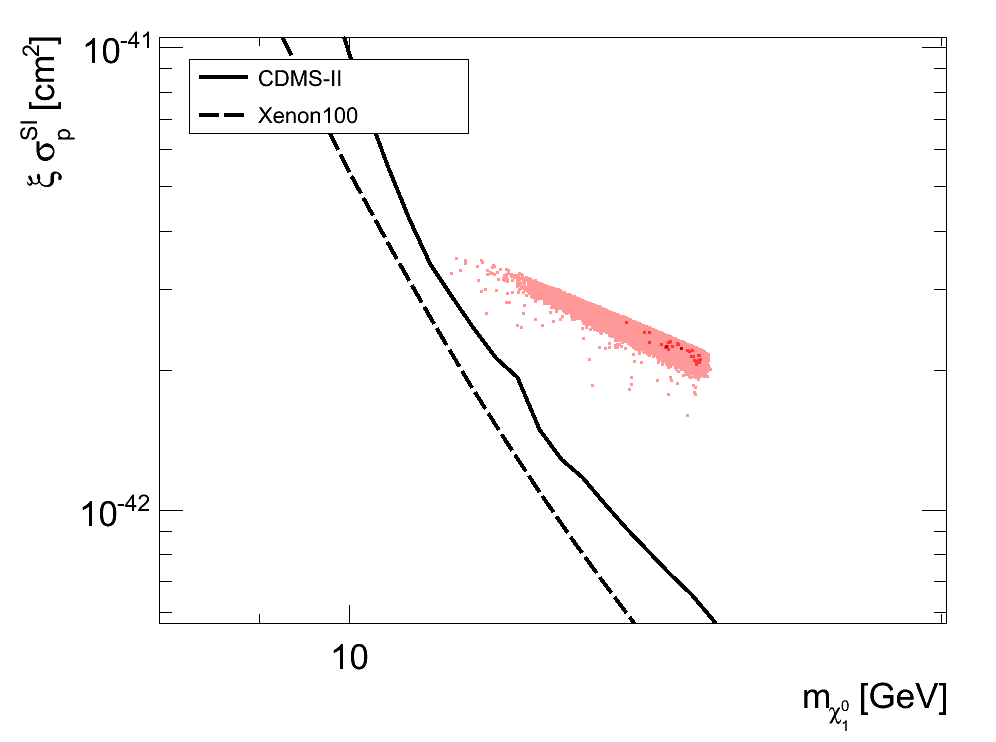}
		\caption{MSSM-EWSB scenario with $\mu>0$ and $m_{\chi} < 15$ GeV. Spin-independent 
		cross section on proton
		times the fraction of neutralinos in the Milky Way dark halo ($\xi$) versus the neutralino mass 
		$m_{\chi}$. The dark red (light pink) points have a likelihood greater than 99.4\% (68\%). }
	\label{fig:SIcsvsMneutralino}
\end{figure}

In Fig.~\ref{fig:SIcsvsMneutralino}, we display the (spin-independent) elastic scattering cross-section 
$\sigma_p^{SI}$ times the fraction of neutralino in the halo $\xi$ versus the neutralino mass and the limits from 
CDMS and XENON 100. Here (and in the following), we have assumed values for the quarks coefficients in the nucleon 
(defined by setting $\sigma_{\pi N}=45 {\rm MeV}, \sigma_0=40{\rm MeV}$ in micrOMEGAs ~\cite{Belanger:2008sj}) 
that lead to rather low  cross-sections in order to be conservative in our predictions 
\footnote{The elastic scattering cross-section can be up to one order of magnitude larger for other choices of the 
quark coefficients.}. Since there are uncertainties on the escape velocity and scintillation function of XENON 100, 
we also performed a rescaling of $L_{eff}$ with the energy  (see \cite{2010PhRvC..81b5808M}) and kept a conservative 
energy-dependent value for $L_{eff}$. In principle, this should enable us to derive the most conservative limit as 
possible.

Since all the light candidates that we have found lie within the excluded region (which is defined by both CDMS 
and XENON 100 exclusion curves), our results strongly suggest light MSSM  neutralinos cannot explain neither
 the CoGeNT nor the DAMA/LIBRA data. Hence, if dark matter is indeed as light as 10 GeV, MSSM neutralinos
  are extremely unlikely to be a viable explanation.

\begin{figure}[h]
	\centering
\includegraphics[width=8cm]{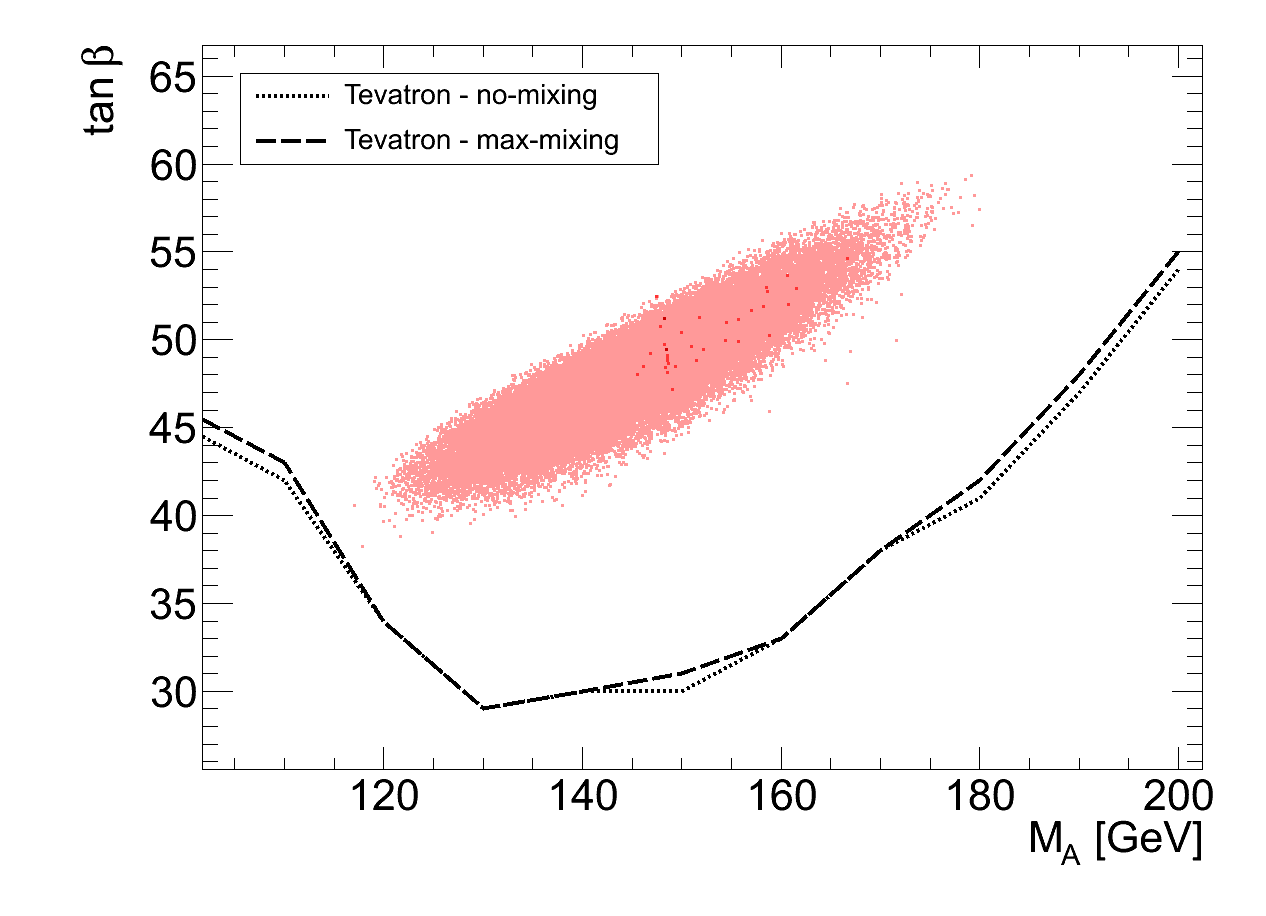}
	\caption{Distribution of the points selected by our MCMC analysis in the $\tan\beta-m_A$ plane in the
	 MSSM-EWSB scenario with $\mu>0$ and $m_{\chi} < 15$ GeV. The TEVATRON limits are displayed for the case of 
	no-mixing (dash) or maximum mixing (full) in the stop sector,
	same color code as in Fig.~\ref{fig:SIcsvsMneutralino}.}
		\label{fig:tanbvsmApos}
\end{figure}

We reach the same conclusion by using collider constraints. Indeed, since the points that we have found are associated with a light pseudo-scalar Higgs, one can use  the $95\%$ exclusion limits from TEVATRON 
in the MSSM plane of $\tan \beta$ versus $m_A$ \cite{Benjamin:2010xb} to constrain the scenarios selected by our MCMC. These collider limits were obtained by 
studying the MSSM Higgs boson production in association with $b$ quarks in the $\tau^+\tau^-$ 
final states with up to 2.2 fb$^{-1}$ of data.  Two benchmark scenarios were in fact considered to describe the situations of maximum and no 
mixing in the stop sector \cite{Carena:2005ek}   for both $\mu>0$ and $\mu<0$. 

In Fig.~\ref{fig:tanbvsmApos}, we thus plot the results of our MCMC analysis and superimpose the TEVATRON constraints. 
We find that, independently of the direct detection constraints,  none of the scenarios corresponding to light 
MSSM neutralinos and $\mu>0$
 can survive the TEVATRON constraints on the mass of the pseudo-scalar Higgs. The fact that neutralinos lighter than 15 GeV ($\mu>0$) are excluded 
 by both direct detection and collider constraints enable us to conclude that MSSM-EWSB neutralinos (in scenarios with $\mu>0$) 
 must be heavier than 15 GeV. This argument is actually valid whatever the precise value of the relic density. This is therefore a very strong conclusion.


Note that our results are somewhat discrepant with \cite{Bottino:2009km}. This may originate from the fact that 
Ref.~\cite{Bottino:2009km} did not take into account the latest constraint on $B_s \rightarrow \mu \mu$. 
Indeed, we do obtain many more points at low mass if we disregard this constraint, as was also pointed out 
in Ref.~\cite{Feldman:2010ke}. However, 
all these ``new'' points  are also excluded by the TEVATRON limits.  

We have performed a similar analysis for the case $\mu<0$. However, 
we found that it is extremely difficult to obtain a good starting point. 
Besides the total Likelihood of the points retained by the 
MCMC is much smaller than that for $\mu>0$, this is mainly due to the $(g-2)_\mu$ constraint and to a lesser extend to 
the constraints from B-physics. 
Hence, MSSM-EWSB neutralinos lighter than 15 GeV are also ruled out in the case $\mu<0$.
\footnote{In Ref.~\cite{Gogoladze:2010fu} it was shown that the $(g-2)_\mu$ constraint could be avoided if one takes
opposite signs for gaugino masses with both $\mu<0$ and $M_2<0$. We have not considered this class of scenarios.}

Of course our lower bound on $\Omega_{\chi} h^2$ (see Eq.\ref{wmap}) is arbitrary. One could consider an even smaller fraction of light neutralinos in the halo.  In this case, not only would they be a minor source of dark matter but also their energy density distribution in the halo may be affected by the nature of the main candidate; this is hard to estimate. 
We have checked nevertheless that taking $\Omega_{\chi} \ll  10 \% \Omega_{WMAP}$ does not change our results. 
This can be understood easily.  
Taking $\Omega_{\chi} \ll \  \Omega_{WMAP}$ requires a large pair annihilation cross-section ($\sigma v_{\chi \chi}$). Yet, $\sigma v_{\chi \chi}$ is proportional to the neutralino mass squared. Hence, the lightest the neutralino, the smallest the pair annihilation cross-section, that is the largest value of $\Omega_{\chi} h^2$. Thus, one cannot take the neutralino mass arbitrarily small. 
To enhance the cross-section, the sole viable option is to invoke neutralino pair annihilations through the exchange of a relatively light pseudo-scalar Higgs (which is in agreement with our findings) since neutralino co-annihilations with the Next-to-LSP 
is impossible owing to the smallness of the neutralino mass.

\subsection{Neutralino masses less than 50 GeV}

Given our conclusion concerning light neutralinos, it is interesting to derive the lower limit for the neutralino mass in the MSSM. For this, we focus on scenarios where the neutralino mass ranges from 1 to 50 GeV. 


\begin{figure}[h]
	\centering		\includegraphics[width=8cm]{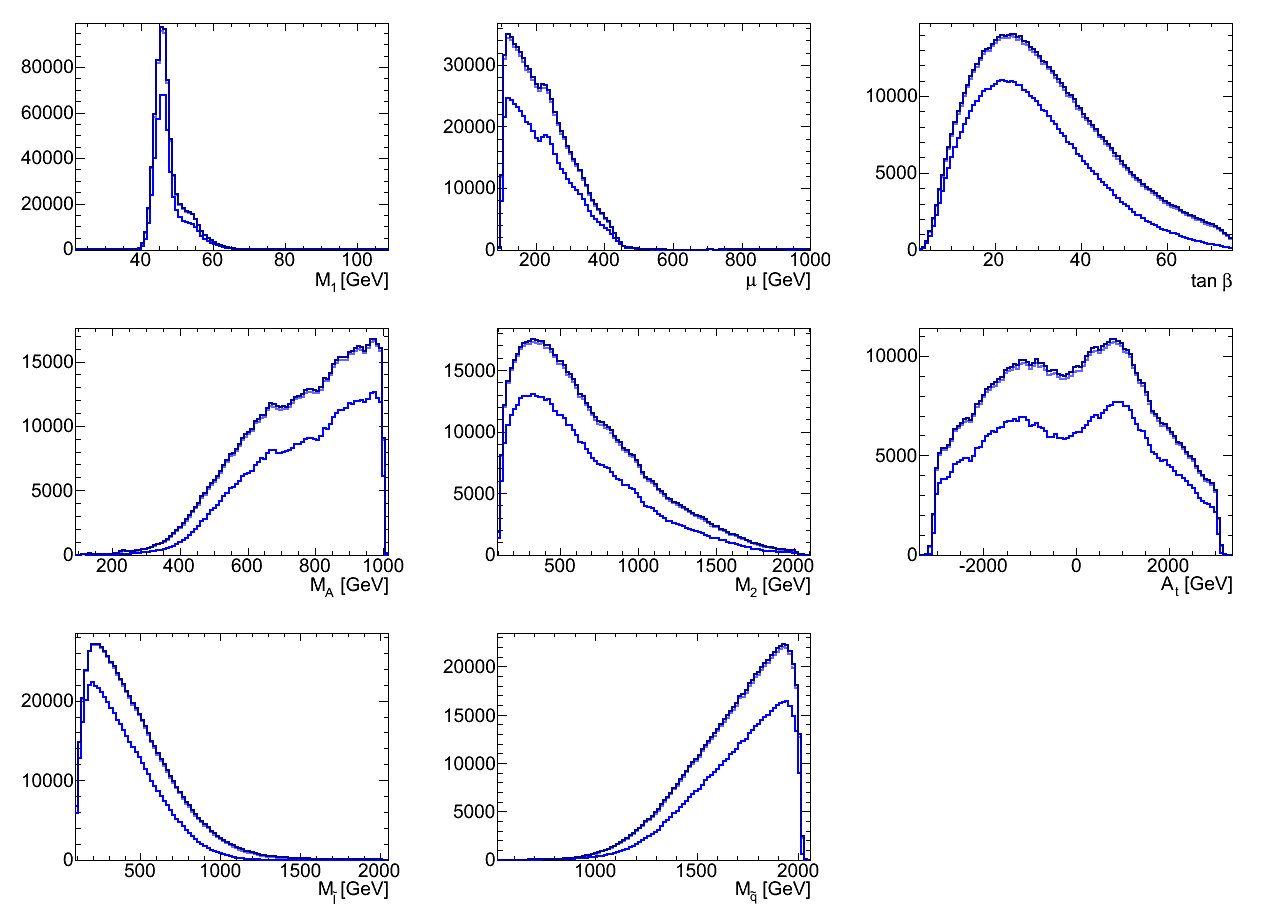}
	\caption{Same as in Fig.~\ref{fig:param} for the MSSM-EWSB scenario with $\mu>0$ and $m_{\chi} < 50$ GeV }.
		\label{fig:paramlarge}
\end{figure}

The distribution of $Q/Q_{max}$ for the free parameters of the model are summarized in
Fig.\ref{fig:paramlarge}.
As one can see, the preferred value of the neutralino mass lies in the [28,50] GeV range and is similar to the value of $M_1$ which lies in $[40,50]$ GeV. Since $M_1 < \mu$, such neutralinos are mostly  bino-like. 
Nevertheless some Higgsino component is necessary for efficient neutralino annihilation through Z or light Higgs exchange, hence the preference  for small values of $\mu$. 
As compared to the previous case, the pseudoscalar Higgs is no longer necessary to provide efficient annihilation, and therefore $m_A$ moves towards higher values. Furthermore $\tan\beta$ varies in a wider range since the $B$-physics constraints are relaxed at large values of $m_A$.

\begin{figure}[h]
	\centering	\includegraphics[width=8cm]{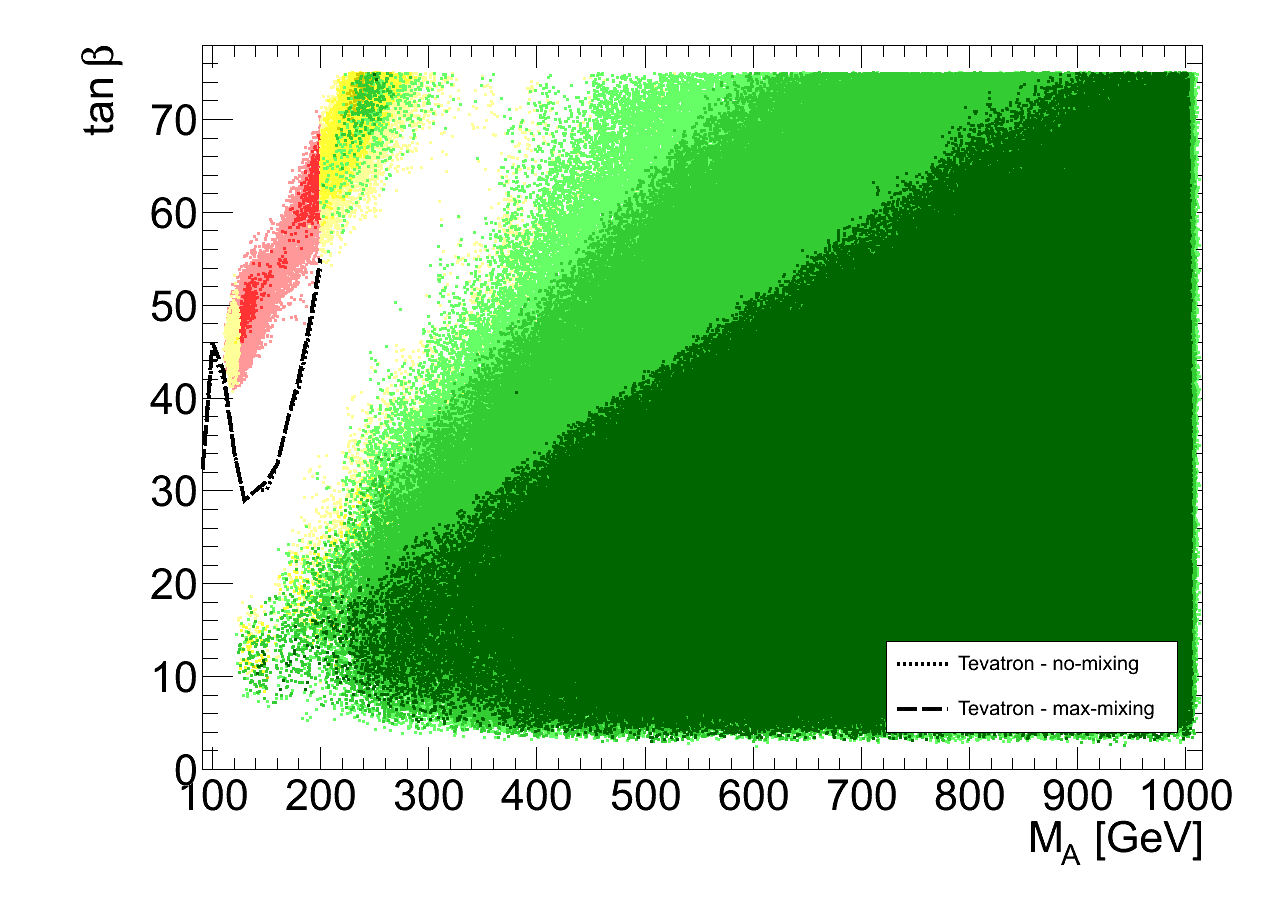}
	\caption{Distribution of the points selected by our MCMC analysis in the $\tan\beta-m_A$ plane in the
	 MSSM-EWSB scenario with $\mu>0$ and $m_{\chi} < 50$ GeV.  In red, we display the points which are 
	excluded by both TEVATRON, XENON 100 and CDMS. In yellow, we show the points which satisfy   TEVATRON and which are 
	excluded by XENON 100 and CDMS and in green, all the points which survive both constraints.}
\label{fig:ewsbhighmasstanma}
\end{figure}

In Fig.~\ref{fig:ewsbhighmasstanma}, we display the points selected by the MCMC  in the plane $(m_{A}, \tan \beta)$ where we have superimposed the TEVATRON constraints. Interestingly enough, at low values of the mass of the pseudo-scalar Higgs $m_A < 300$ GeV, there are two separate regions of $\tan \beta$. One is peaked around 10-20 while the second lies between $50-70$. Typically when the  pseudoscalar is light, constraints on $B$-physics decrease the value of the likelihood especially when $\tan\beta$ is large. However as we have seen previously the new channel for neutralino annihilation through a pseudoscalar exchange leads to an acceptable relic density and to a good global likelihood when $\tan\beta > 50$. Note that, even though very large values of $\tan \beta$ do not appear plausible, they do indicate the type of regions that
 lead to a neutralino mass in the $[1,50]$ GeV range. 

\begin{figure}[h]
	\centering \includegraphics[width=8cm]{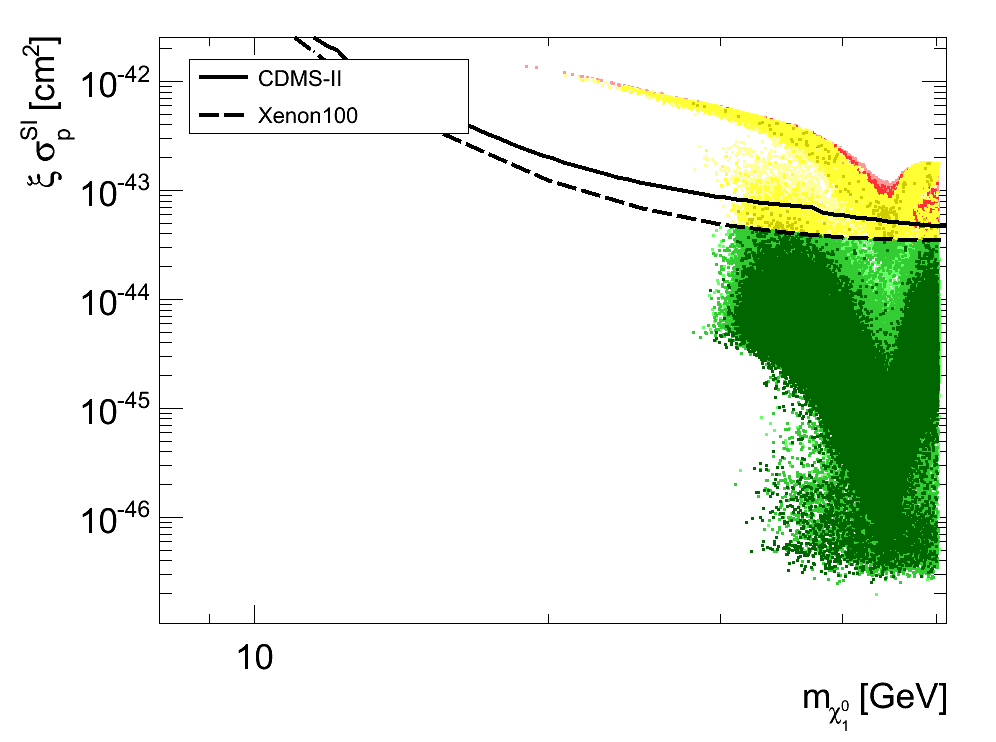}
	\caption{Spin independent cross section vs the neutralino mass in the MSSM-EWSB scenario with $\mu>0$ and $m_{\chi} < 50$ GeV, 
	same color code as in Fig.~\ref{fig:ewsbhighmasstanma}}
		\label{fig:sicslarge}
\end{figure}

The spin-independent cross section versus the neutralino mass is displayed in Fig.~\ref{fig:sicslarge}. 
The scenarios where the neutralino mass is greater than 28 GeV survive both the TEVATRON and Direct Detection limits. 
Although such a value is likely to be irrelevant to explain CoGeNT data, it might be important in light of the two CDMS ``events''.

For $\mu<0$, we again found that the Likelihood of the points is much smaller than for $\mu>0$, owing mainly to the 
$(g-2)_\mu$ constraint which disfavours large values of $\tan \beta$ as well as low values of $M_2$ and 
$m_{\tilde l}$ ~\cite{Belanger:2003wb}.


\section{NMSSM scenarios}

The Next-to-Minimal Supersymmetric Standard Model (NMSSM) is a simple extension of the
MSSM that provides a solution to the naturalness problem. This is achieved
by the introduction of a gauge singlet superfield, denoted by S. The VEV of this singlet  determines the effective
parameter $\mu=\lambda \langle S \rangle$ which is then naturally of the EW scale ~\cite{Ellwanger:2009dp}. 
The part of the superpotential involving Higgs fiels reads
\begin{equation}
 W=\lambda S H_uH_d +\frac{1}{3} \kappa S^3 
 \end{equation}
and  the  soft Lagrangian 
\begin{eqnarray}
{\cal L}_{\rm soft} = m^2_{H_u}|H_u|^2 + m^2_{H_d}|H_d|^2
+m^2_{S}|S|^2\ \nonumber\\
+(\lambda A_\lambda H_u H_d S+ \frac{1}{3} \kappa A_\kappa S^3+h.c.)
\end{eqnarray}
The NMSSM contains three neutral scalar fields, $h_1,h_2,h_3$ and two pseudoscalar neutral fields, $a_1,a_2$ as well as a charged Higgs, $H^\pm$.
The model also contains five neutralinos, the new field is the singlino, $\tilde S$. For a pure state the
singlino mass is simply ~\cite{Ellwanger:2009dp}
\begin{equation}
m_{\tilde S} = 2 \frac{\kappa\mu}{\lambda}.
\label{eq:ms}
\end{equation} 
After using the minimization conditions of the Higgs potential, the Higgs sector is described by
six free parameters, $\mu,\tan\beta$ as well as $\lambda, \kappa, A_\lambda, A_\kappa$.
Other free parameters of the model are, as in the MSSM, the soft masses for sfermions, 
trilinear couplings and gaugino masses.
 
An important feature of the model is that both the singlino and the singlet fields can be very light and yet escape the LEP bounds. 
This is because these fields mostly decouple from the SM fields ~\cite{Ellwanger:2009dp}.
This opens up the possibility for new annihilation mechanisms for light neutralinos in particular if the LSP possesses
 an important singlino component. 
The singlino can annihilate efficiently through the exchange of 
light singlet Higgses as well as into light Higgs singlets~\cite{Belanger:2005kh}.

\subsection{Neutralino masses smaller than 15 GeV}
To explore the parameter space of the NMSSM model that allow for light neutralinos we follow the same procedure 
as for the MSSM. Our priors lie in the range:

\begin{equation}
\begin{array}{cc}
M_1 \in [1,200] {\rm GeV}  & M_2 \in [100, 2000]{\rm GeV} \nonumber\\
\mu \in [0.,1000] {\rm GeV}  &\tan \beta \in  [0.1,65]  \nonumber\\
\lambda \in [0,0.75] & \kappa\in [0.,0.65] \nonumber\\
A_\lambda \in [-2000,5000]{\rm GeV} & A_\kappa \in [-5000,2000]{\rm GeV} \nonumber\\
m_{\tilde{l}} \in [100,2000]{\rm GeV} & m_{\tilde{q}} \in  [300,2000]{\rm GeV} \nonumber\\
A_t \in [-3000,3000]{\rm GeV} & 
\end{array}
\end{equation}

As before, we assume common soft masses for 
squarks and sleptons and we keep the gaugino masses $M_1$ and $M_2$ uncorrelated while $M_3 = 3 M_2$.

\begin{figure*}
	\centering\includegraphics[width=15cm]{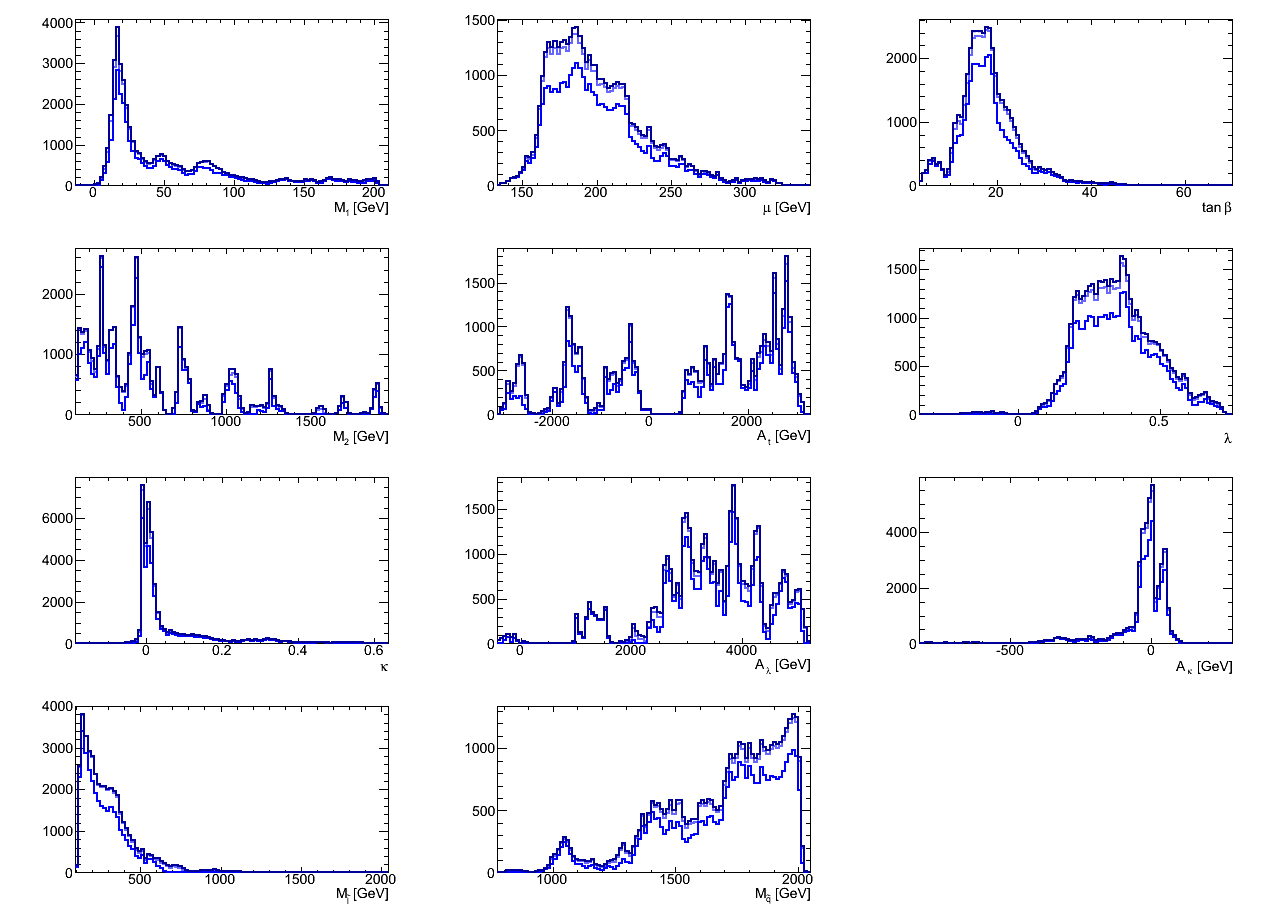} 
	\caption{$Q/Q_{max}$ for the points selected by 
	the MCMC versus the value of the free 
	parameters that we have considered in the NMSSM for $\mu>0$ and $m_{\chi} < 15$ GeV, same color code as Fig.1} 
		\label{fig:param_nmssm}
\end{figure*}

The distribution of  $Q/Q_{max}$ for the free parameters are displayed in Fig.~\ref{fig:param_nmssm}. As in the MSSM the bino mass is 
peaked below 20GeV although a long tail extends to 200GeV. In this tail the LSP is mostly a singlino which mass  
is determined from Eq.~\ref{eq:ms}. 
The parameter $\lambda$ that determines the mixing of the singlino to other neutralinos is never very small, 
so that the singlino does not decouple completely. 
The preferred values for $\mu \approx 150-250$~GeV are higher than in the MSSM.  On the one hand
LEP2 limits on $e^+e^-\ra \tilde\chi^0_1\tilde\chi^0_i$ or on the light Higgs constrain low values of $\mu$ 
while a light singlino LSP prefers low values for $\mu$, Eq.~\ref{eq:ms}.
The parameter $\kappa << 1$ also favours a light singlino. 
Intermediate values of $\tan\beta$ are preferred.  
The parameter $A_\kappa$ that controls
the mass of the singlet Higgses is always small to ensure a  light scalar/pseudoscalar as required for LSP annihilation 
while $A_\lambda$ is usually well above 1 TeV.
Sleptons are preferably light while squarks are above 1TeV.

\begin{figure}[htb]
	\centering		\includegraphics[width=8cm]{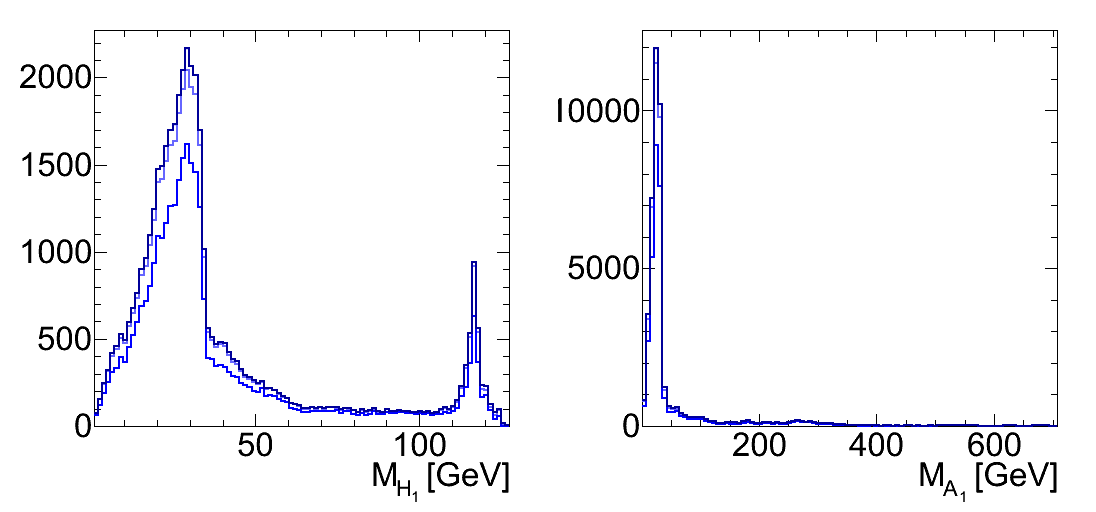}
	\caption{$Q/Q_{max}$ with respect to  the mass of the  scalar (left)  and pseudoscalar Higgs (right)
	in the NMSSM  for $\mu>0$ and $m_{\chi} < 15$ GeV, same color code 
	 as in Fig.~\ref{fig:param}.}
	\label{fig:higgs_nmssm}
\end{figure}

\begin{figure}[htb]
	\centering		\includegraphics[width=8cm]{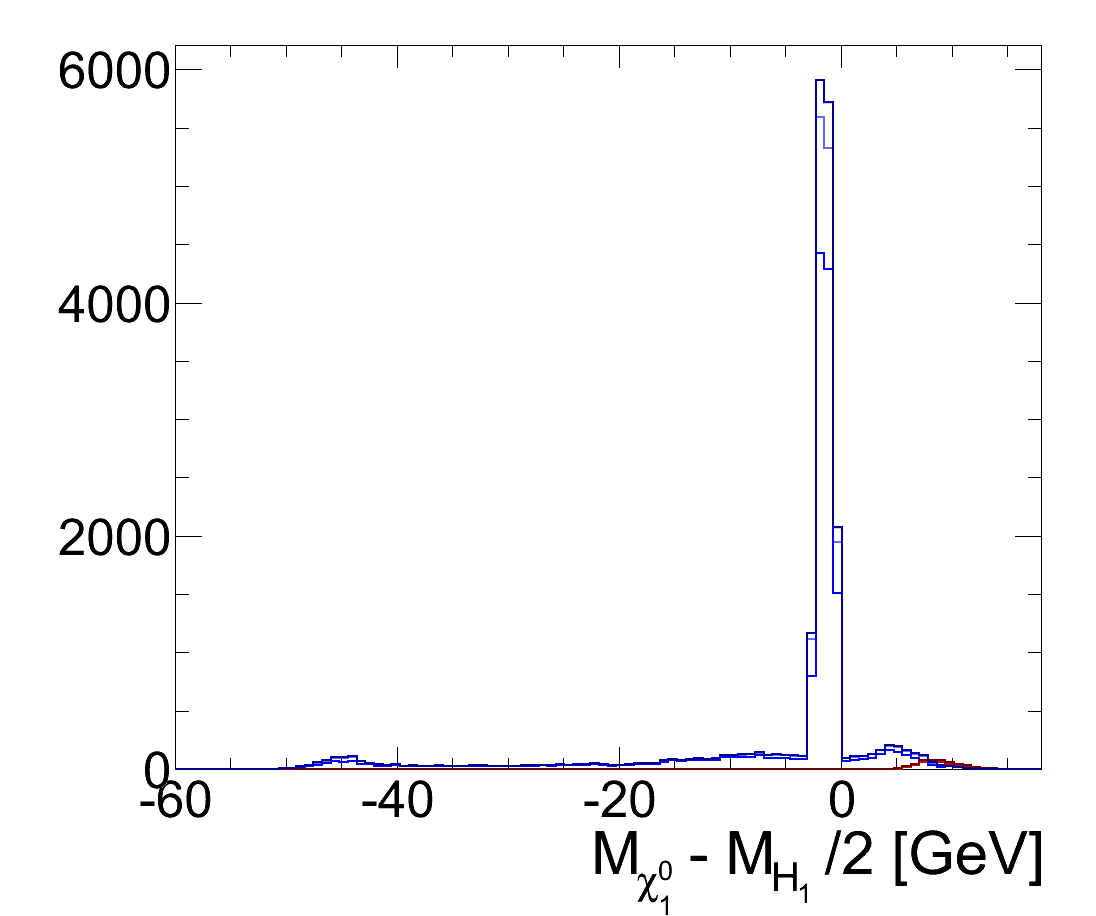}
	\caption{NMSSM scenario for $\mu>0$ and $m_{\chi} < 15$ GeV.  Mass difference $m_{\chi}-m_{a1}$
	We use the same color code as in Fig.~\ref{fig:param}.}
	\label{fig:mass_splitting}
\end{figure}

The LSP mass ranges from 1-15GeV with a distribution peaked towards higher masses. This LSP is either mostly bino or mostly 
singlino with in any case some higgsino component.   
The most important feature of this scenario is the fact that the Higgs spectrum is constrained: one always predict a light scalar, dominantly singlet,
with a mass  below 120 GeV (generally below 30GeV) as well as a pseudoscalar singlet with a mass preferably below 30GeV, 
see fig.~\ref{fig:higgs_nmssm}. Note that the value of 30GeV for the mass corresponds to twice the neutralino mass and is thus just a consequence of the prior on the neutralino mass.
Furthermore we find generally that either $m_{\lsp}-m_{a_1}/2 <1-4 ~{\rm  GeV}$ (with a similar mass splitting with $h_1$, cf
Fig.~\ref{fig:mass_splitting}) or that  
$m_{\lsp}> m_{h_1}$.
This is because the annihilation of the light LSP relies either on pseudoscalar/scalar exchange or on the new light scalar pairs final states. 
The rest of the Higgs sector consists of MSSM-like doublets with preferred values for the heavy neutral and charged scalars
above 2TeV. Note that we have checked that the recent re-analysis of LEP2 limits on a Higgs decaying into two light pseudo-scalars did not put further constraints on
our model parameters ~\cite{Schael:2010aw}.

The light LSP scenarios can be classified  in three broad classes: 
1) a (pure or mixed) singlino LSP annihilating via 
pseudoscalar/scalar singlet Higgses into fermion final states, for this only a small singlino component of the LSP is necessary.
2) a bino LSP  with  small higgsino/singlino components annihilating into a pair of light scalar Higgses
or 3) as in the MSSM a bino LSP with some Higgsino component annihilating via Higgs doublets.
This channel is more efficient  at large values of $\tan\beta$ although the B-physics constraints severely restrict the very large values of $\tan\beta$.

\begin{figure}[htb]
	\centering
\includegraphics[width=8cm]{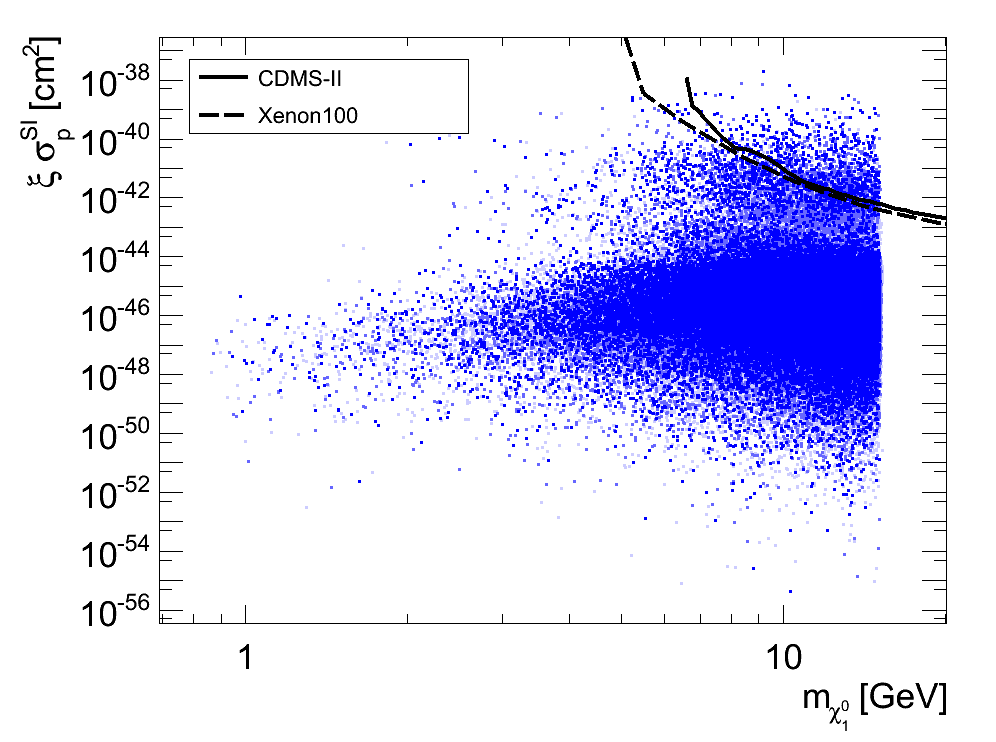}
		\caption{NMSSM scenario with $\mu>0$ and $m_{\chi} < 15$ GeV. 
		Spin-independent neutralino-proton cross section times the fraction of neutralinos in the Milky Way dark halo 
		($\xi$) versus the neutralino mass $m_{\chi}$. 
		We use the same color code as in Fig.~\ref{fig:param}. The CDMS limits correspond to 
		the plain curve while the XENON limits correspond to the dash curve.}
	\label{fig:SIcsvsM_nmssm}
\end{figure}

The predictions for the  elastic scattering cross section span several orders of magnitude, from $10^{-56}$ to  $10^{-38}{\rm cm}^2$, see Fig.~\ref{fig:SIcsvsM_nmssm}.
The largest cross sections are found in scenarios with a light $h_1$, 
for example for $\sigma^{SI}_{\chi p}>10^{-43}(10^{-41}){\rm cm}^2$ requires $m_{h_1}<20(8)$~GeV.
\footnote{In Ref.~\cite{Bae:2010hr}, light  CP-even Higgs exchange was proposed to explain the CoGeNT result 
with  light neutralinos within the BMSSM model.}
At first sight this can be a bit surprising since such a light Higgs is dominantly singlet and thus couples very weakly to quarks in the nucleon - recall
that the $h_1 q\bar{q}$ coupling is only possible through the doublet component- 
nevertheless this  suppressed coupling is compensated by an enhancement factor due to the small $h_1$ mass in the propagator,
 $\propto 1/m_{h_1}^2$.
 In scenarios where the elastic scattering cross-section is large, the LSP is generally dominantly bino (or, in a few cases, a singlino) with a non-negligible 
higgsino fraction. This means that the doublet $h_2$ also contributes to the   
spin independent cross section since the LSP coupling to the doublet depends on
the Higgsino component of the LSP.
\footnote{Note that in Ref~\cite{Das:2010ww} an upper bound on the SI cross-section
was obtained by optimizing the contribution of the doublet exchange, we found larger cross sections
 because we explored regions where the additional contribution of the light singlet was important.}
The very low cross-sections are found in  scenarios where the LSP pair-annihilation benefits from the  enhancement 
of the  pseudoscalar exchange in the s-channel near the resonance  
while the elastic scattering cross-section, which proceeds through scalar exchange in t-channel, 
does not benefit from a similar enhancement. 

\begin{figure}[h]
	\centering		\includegraphics[width=8cm]{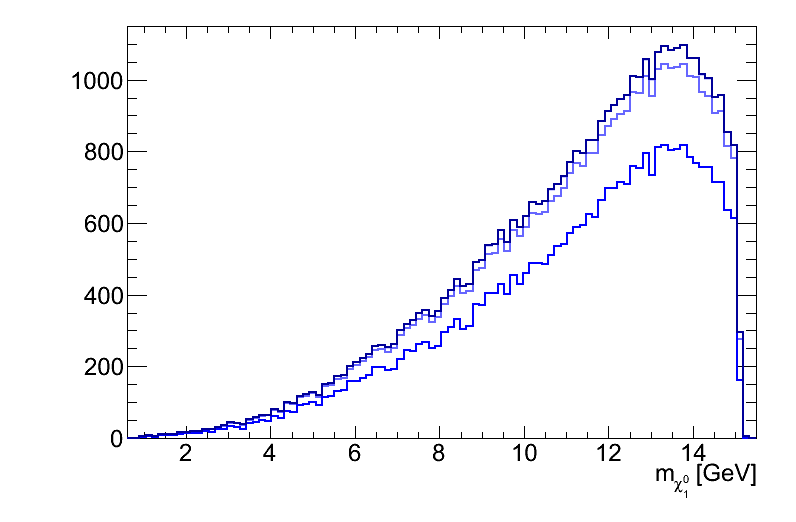}
	\caption{$Q/Q_{max}$ with respect to  the neutralino mass in the NMSSM  for $\mu>0$ and $m_{\chi} < 15$ GeV, 
	same color code as in Fig.~\ref{fig:param}.}
	\label{fig:distrib_nmssm}
\end{figure}

We have also explored the region of parameter space which gives neutralino masses up to 50GeV. 
As for the MSSM we found many scenarios that satisfies all the constraints. 
These have a neutralino mass near $M_Z/2$ and a SI cross section below the
present limits.  Note that we also found that neutralinos lighter than 15 GeV were accompanied by a singlet 
pseudoscalar having a mass $\approx 2m_\chi$. However, these scenarios did not have a significant singlino 
component. In addition, they had a lower likelihood than in the case of heavier neutralinos.

\subsection{LHC signatures for light neutralino scenarios}

Light NMSSM neutralino scenarios ($m_{\chi}<15$GeV) 
can lead to distinctive signatures at colliders both in the Higgs and the neutralino sector. 
First, the usual dominant decay mode of the light  scalar doublet Higgs can be greatly suppressed
because of new decay modes into neutralinos $h_2\rightarrow\lsp\lsp$ 
as well as into light Higgses $h_2\rightarrow h_1 h_1,a_1 a_1$.  
The light scalar/pseudoscalar are expected to decay predominantly into $b\bar{b},\tau^+\tau^-$ or invisibly
into $\lsp\lsp$. Therefore both the search for the invisible Higgs 
in the W fusion channel with a signature in two tagged jets and missing $E_T$ 
and the search for Higgs via  $Wh_2$ production  with $h_2\rightarrow h_1 h_1\rightarrow 4j$
~\cite{Cheung:2007sva,Carena:2007jk} are important channels at the LHC.

Distinctive signatures are also expected in the neutralino and chargino sectors. The NLSP, $\tilde\chi^0_2$,
can contain a small component of singlino and  therefore  be much lighter than $\tilde\chi_1^+$.
The favourite discovery channel, $\tilde{\chi}^0_2\rightarrow l\bar{l}$ is however much suppressed since
the decay modes can involve light Higgs states,  $\tilde\chi^0_2\rightarrow \lsp h_1,\lsp a_1$, leading to 
$\tilde\chi^0_2\rightarrow \lsp b\bar{b}$  or to the completely invisible decay 
$\tilde\chi^0_2\rightarrow \lsp \lsp \lsp$.  
Furthermore 
the single lepton signature of the chargino will be much suppressed since  
the decay $\tilde\chi^+\rightarrow W^+\tilde\chi_1^0$  is generally accessible because of the large mass 
splitting with the LSP.

\section{Conclusion}

We have investigated whether light neutralinos in the MSSM and the NMSSM could still be viable candidates and furthermore
lead to large values of the elastic scattering cross-section as required to explain recent CDMS, 
CoGeNT, DAMA/LIBRA  data. 

In the MSSM, we basically exclude neutralinos lighter than 15 GeV. In order to satisfy the WMAP constraint, these particles must exchange a light pseudoscalar Higgs at large values of $\tan\beta$. However, recent TEVATRON results on supersymmetric Higgs searches rule out this possibility. In addition, we found (after imposing all collider and direct detection constraints) that the lower limit on the neutralino mass is about 28 GeV in the MSSM ($\mu>0$).  

In the NMSSM, it is  easier to find light neutralinos that satisfy all the constraints. 
This is generally achieved when the LSP contains a singlino component. One salient feature is that these neutralinos
are accompanied by a light pseudoscalar and/or scalar singlet. 
While the elastic scattering cross-sections are generally much below the reach of current detectors in most scenarios, 
we found that some points can satisfy all the constraints and yet predict elastic scattering cross-sections in the ``CoGeNT'' region. This is achieved  through the presence of a 
${\cal O} ({\rm GeV})$ scalar Higgs. However we should stress that the light neutralinos typically do not provide as good a
 fit to the data as the ones around 50~GeV.

NMSSM scenarios with light neutralinos could  be further probed 
with direct detection experiments of increased sensitivity. They also 
lead to distinctive signatures in  Higgs and SUSY particles searches at colliders that could be studied. 
Finally, indirect signatures of the light neutralino (including synchrotron emission, see Ref.~\cite{Boehm:2010kg}) will be investigated in a future work.

 {\bf Note added}: After submitting this paper, we became aware of another
 preprint ~\cite{Draper:2010ew} which also finds
 that light neutralinos in the NMSSM can have large elastic scattering cross sections through
 the exchange of a light scalar Higgs.

\section*{Acknowledgment}
We would like to thank A. Nikitenko for providing useful references on Higgs limits and C. Hugonie for discussions. 
This work is  supported in part by the GDRI-ACPP and PICS of CNRS and by the French 
ANR project {\tt ToolsDMColl}, BLAN07-2-194882.
The work of AP is supported by the Russian foundation for Basic Research, 
grant RFBR-08-02-00856-a and RFBR-08-02-92499-a. 
D. A. is supported by a grant ``EXPLORADOC'' and is thankful to S. Sarkar for
supporting his application.  
D.A. and C.B. would like to thank the Astrophysics department of the 
University of Oxford for hospitality. 
\bibliography{ewsb}
\end{document}